\documentclass[%
 reprint,
 amsmath,amssymb,
 aps,
superscriptaddress]{revtex4-2}

\usepackage{graphicx}
\usepackage{dcolumn}
\usepackage{bm}
\let\PhysLett\pl
\let\pl\relax
\usepackage{unitsdef}
\let\pl\PhysLett

\usepackage{xcolor}
\usepackage{amsmath}
\usepackage[english]{babel}
\makeatletter
\def\l@en{\l@english}
\makeatother

\makeatletter
\def\l@eng{\l@english}
\makeatother

\begin{document}

\preprint{APS/123-QED}

\title{A Superconducting Phase Transition Single-Electron Transistor}

\author{Gorka Aizpurua-Iraola}
\affiliation{Quantum Motion, 9 Sterling Way, London, N7 9HJ, United Kingdom}
\affiliation{CIC nanoGUNE Consolider, Tolosa Hiribidea 76, E-20018 Donostia-San Sebastian, Spain}
\author{Thomas H. Swift}
\email{tom@quantummotion.tech}
\affiliation{Quantum Motion, 9 Sterling Way, London, N7 9HJ, United Kingdom}
\author{Felix-Ekkehard von Horstig}
\affiliation{Quantum Motion, 9 Sterling Way, London, N7 9HJ, United Kingdom}
\author{Domenic Prete}
\affiliation{Quantum Motion, 9 Sterling Way, London, N7 9HJ, United Kingdom}
\author{James Kirkman}
\affiliation{Quantum Motion, 9 Sterling Way, London, N7 9HJ, United Kingdom}
\author{Grayson M. Noah}
\affiliation{Quantum Motion, 9 Sterling Way, London, N7 9HJ, United Kingdom}
\author{Fabio Olivieri}
\affiliation{Quantum Motion, 9 Sterling Way, London, N7 9HJ, United Kingdom}
\author{Alberto~Gomez-Saiz}
\affiliation{Quantum Motion, 9 Sterling Way, London, N7 9HJ, United Kingdom}
\affiliation{Department of Electrical and Electronic Engineering, Imperial College London, London SW7 2AZ, United Kingdom}
\author{M. Fernando Gonzalez-Zalba}
\email{fernando@quantummotion.tech}
\affiliation{Quantum Motion, 9 Sterling Way, London, N7 9HJ, United Kingdom}
\affiliation{CIC nanoGUNE Consolider, Tolosa Hiribidea 76, E-20018 Donostia-San Sebastian, Spain}
\affiliation{IKERBASQUE, Basque Foundation for Science, E-48011 Bilbao, Spain}

\date{\today}

\begin{abstract}

Quantum computers require fast and accurate methods for qubit state detection. Phase-transition sensors exploit the abrupt change between two physical states of a material to achieve enhanced sensitivity and have enabled advanced detectors for quantum technologies, such as superconducting nanowire single-photon detectors. However, this sensing principle has not yet been applied to semiconductor spin qubits. Here, we demonstrate a superconducting phase-transition radio-frequency single-electron transistor (PTSET), a charge sensor for semiconductor spin qubits whose response is enhanced by a superconducting-to-normal phase transition. The transition is engineered by linking the sensor current to a low-critical-current, high-kinetic-inductance inductor integrated into the radio-frequency matching network. We demonstrate improvements in sensitivity of one and two orders of magnitude over conventional rfSETs in the large- and small-signal regimes, respectively. Our results establish phase-transition sensing as a route towards ultrasensitive, integrated charge sensors for semiconductor quantum computing and point to broader applications, including cryogenic photon detection for radio astronomy.

\end{abstract}

\maketitle 

\section{Introduction}

Phase transition sensors are a class of detectors in which the detection event is associated with a transition between two physical states of a material. The transition can be superconducting/resistive, insulating/metallic, magnetic/non-magnetic and structural~\cite{Goltsman2001, IrwinHilton2005}. Their fundamental attraction is that they do not necessitate the microscopic excitation itself to be large. Instead, the detector can be engineered close to a transition, where a small perturbation produces a disproportionately large change in a macroscopic property~\cite{DAndrea2024}. Phase transition detectors are already important in quantum computing with examples such as superconducting nanowires for photonic implementations~\cite{You2020}. 

Spins in semiconductor quantum dots are a particularly promising platform for large-scale integration of quantum computers, due to their proven high control and readout fidelities~\cite{wu2026simultaneoushighfidelitysinglequbitgates, Xue2022QuantumThreshold, Laine2026}, their compatibility with semiconductor manufacturing~\cite{Reilly2019ChallengesComputer, Xue2021CMOS-basedCircuits, Ruffino2022AElectronics} and the possibility of all-digital cryogenic addressing~\cite{Abraham2026}. However, spin readout remains the slower step, ultimately limiting the time scale of error correction cycles~\cite{Undseth2026}. Finding ways to improve current sensing methodologies through phase transition detection could provide a useful path to this target.  

The radio-frequency single-electron transistor (rfSET) is one of the most common approaches to spin qubit readout since it can achieve single electron charge sensitivity at sub-microsecond measurement rates~\cite{RF_SET_Schoelkopf,PhysRevB.81.161308,PhysRevLett.86.3376, Keith2019}. The rf-SET consists of two main elements: (i) a charge sensor and (ii) an impedance matching network. The charge sensor comprises an isolated nanoscale island (which can be metallic, semiconducting, or superconducting) arranged in a three terminal configuration (source, drain, and gate). Charge sensing occurs by measuring a source-drain tunnelling current through the island at a fixed gate bias set on the slope of a Coulomb blockade oscillation (CBO), where the current exhibits a strong dependence on the surrounding electrostatic potential. The addition or removal of charge in a neighbouring structure perturbs the local electric field, thereby shifting the CBO along the gate-voltage axis and producing a corresponding change in current. 

The impedance matching allows overcoming the measurement bandwidth limitations of the SET caused by its large source-drain resistance (of at least twice the resistance quantum, $R_\text{SET}\geq 2R_Q$). The network inverts the characteristic impedance towards 50~$\Omega$ allowing fast impedance reflectometry measurements through low-noise cryogenic amplification~\cite{10.1063/5.0088229}. This network consists of an $LC$ circuit in which the geometrical inductance of either a normal metal or a superconductor is utilized -- the superconducting approach providing lower loss and hence larger impedance contrast between the different sensing states of the SET.

In this work, we present a novel approach to rf-SET utilizing, specifically, a low critical current, high-kinetic inductance superconductor as the inductive element. By tuning the critical current below the maximum tunnelling current through the SET, and allowing the current to flow in series through the inductor and the SET island, a change in current produced by a charge sensing event triggers a superconducting-to-normal phase transition making the inductor become a normal metal resistor. This transition produces a sudden change in the circuit impedance, significantly enhancing the sensitivity of the sensor. We call this rf-SET variant the superconducting phase transition SET (PTSET).



\section{Superconducting Phase Transition SET proposal}

\begin{figure}[ht!]
\includegraphics[width=\columnwidth]{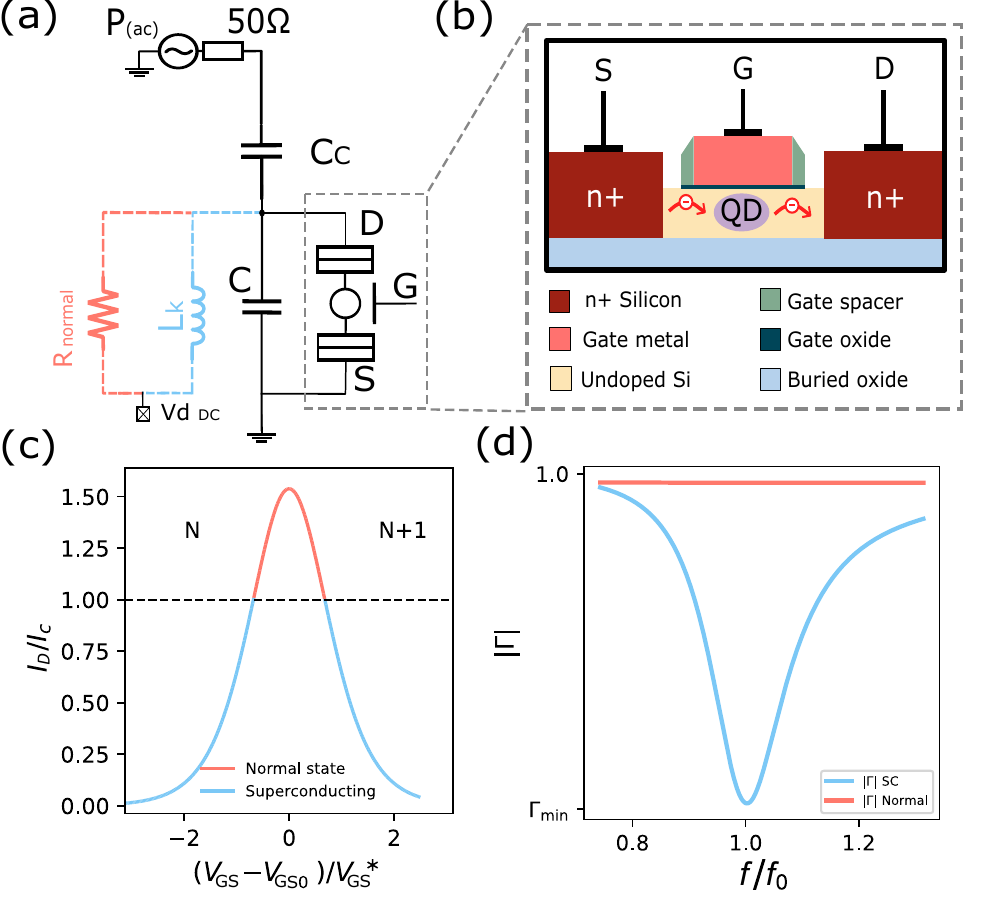}
\caption{\label{fig:1}\textbf{PTSET principle of operation.} (a) Schematic view of the matching circuit with both TiN states: resistive state (red), superconducting state (blue). (b) Schematic cross-section of the SET showing a charged island at the silicon layer directly under the gate as well as the flow of the single-electron AC current (red arrows). (c) Simulated Coulomb oscillation: Normalized drain current to the critical current of the superconductor thin film vs normalized gate-source voltage. (d) Calculated reflection coefficient |$\Gamma$| as a function of the frequency (normalized to the resonance frequency, $f_0$) for both TiN thin film possible states: Superconducting state (blue) and resistive state (red).}
\end{figure}

We present the concept of the PTSET in Fig.~\ref{fig:1}. In panel a, we show a schematic of the matching network used in this experiment consisting of a high-pass low-high network variant in which the SET is embedded (see panel b for schematic cross-section of the SET, whose specific physical implementation is described in detail later). The coloured part of the circuit schematic shows the lumped element representation of the two different electrical states of the high kinetic inductance superconductor. Below the critical current, the element behaves as a superconducting inductor which we model as a purely inductive component, $L_K$ (path in blue). Above the critical current, the same element is modelled as a pure normal resistor $R_\text{norm}$ (path in red).  Critically, we link the state of the superconducting element to that of the SET by the topology of our circuit, allowing the current to flow through the superconductor and the SET from the dc port D of the resonator to the source of the device (connected to ground). The configuration triggers a phase transition once the current through the SET ($I_D$) surpasses the critical current of the superconductor $I_c$ (see Fig~\ref{fig:1}(c)).

This phase transition translates into an abrupt change in the resonator impedance (see panel (d)), effectively producing a large charge in the reflection coefficient $\Gamma=(Z-Z_0)/(Z+Z_0)$ where $Z$ is the impedance of the network (including the SET) and $Z_0$ is the characteristic impedance~\cite{pozar2012microwave}. Specifically, the change in the reflection coefficient ($\Delta\mathrm{\Gamma}$) can be expressed as:


\begin{equation}
    \Delta \Gamma \rightarrow \frac{2\beta}{1+\beta},
    \label{eq:delta_gamma_vs_beta}
\end{equation}

\noindent where $\beta=Z_0/Z$ is the coupling coefficient in the superconducting state. Panel (d) illustrates the abrupt change in reflection at the resonant frequency ($f_0$), rising from $\Gamma_\text{min}= (1\mp \beta)/(1 \pm \beta)$ $(\text{for } \beta \gtrless 1)$) to a state where nearly the entire signal is reflected ($\Gamma\rightarrow1$). 





To implement this concept experimentally, we design an integrated circuit using the GlobalFoundries 22-nm fully-depleted silicon-on-insulator CMOS process~\cite{Ong2017_22FDX} and measure it at the base temperature of a dilution refrigerator. The circuit consists of a TiN superconducting thin film featuring a 1.1 K critical temperature, a high kinetic inductance (122 nH at zero magnetic field) and a low critical current which can be tuned below the SET tunnelling current, as we demonstrate later. The film is part of a polycrystalline resistor stack with a nominal width of 360~nm, a length of 50~$\mu$m and an estimated thickness of 3~nm~\cite{Swift:2025wkk}. The SET is implemented in a minimum width (80~nm) and 28-nm gate length field-effect transistor (FET) showing Coulomb blockade in the sub-threshold region~\cite{DeKruijf2024,Thomas2025}. The matching network is completed by metal-oxide-metal (MOM) capacitors for line coupling ($C_C = 164$~fF) and for the resonator ($C = 114$~fF).


\section{\label{sec:1} Tuning into Operation Mode}

\begin{figure}[ht!]
\includegraphics[width=\columnwidth]{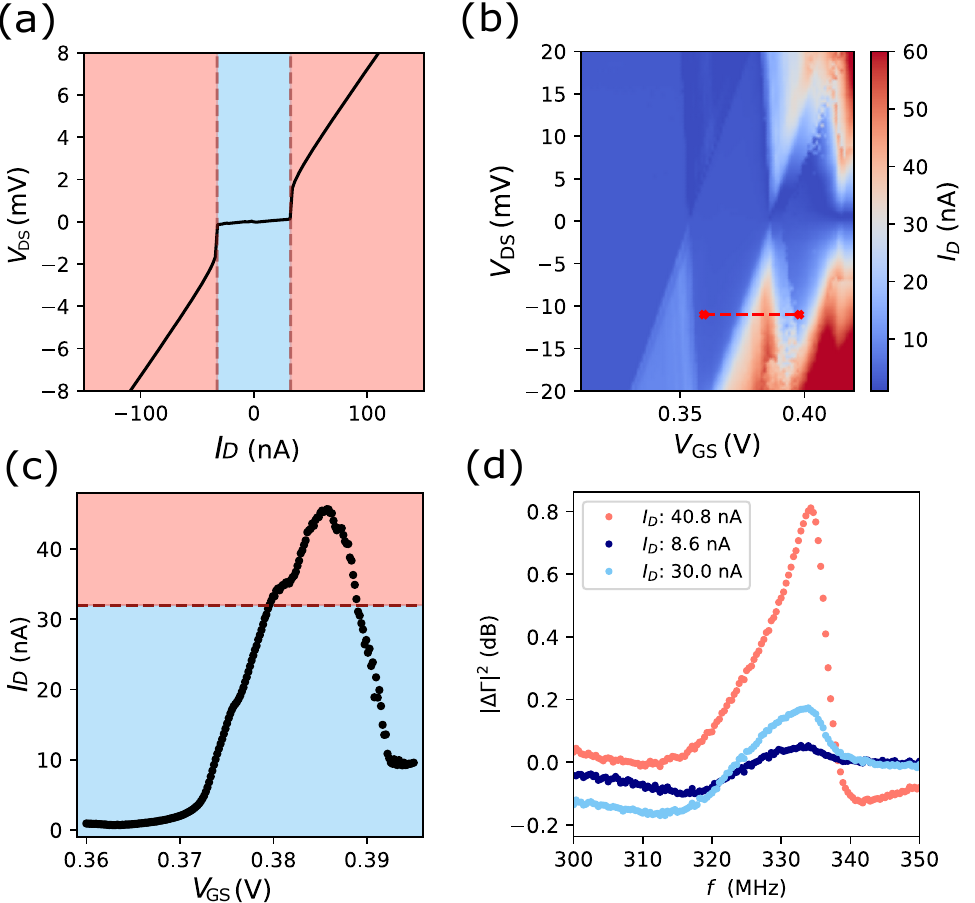}\caption{\label{fig:2} \textbf{Circuit tuning of the PTSET.} (a) Current-biased IV curve measurements of the TiN thin films taken at 0.6 T out-of plane field and $V_{GS}=1$~V, well in the on state of the transistor. A constant slope of 312~k$\Omega$, 112~k$\Omega$ associated with the on-state resistance of the FET and 200~k$\Omega$ with drain and source low-pass filters, is removed to highlight the superconducting region. Red and blue backgrounds represent normal and superconducting states respectively. Vertical dashed lines indicate the phase transition current. (b) Drain current as a function of the gate-source and drain-source voltages, showing regions of charge stability, i.e. Coulomb diamonds. The horizontal red dashed line marks the voltage region used for panel c. No magnetic field applied. (c) Drain current versus the gate-source voltage for a Coulomb oscillation meeting the phase transition tune condition. Red and blue backgrounds represent normal and superconducting states, respectively. The horizontal dashed line indicates the phase transition current. (d) Change in reflected rf power versus frequency for different drain currents performed at 0.6 T out-of plane field. The background is taken at $I_D=0$~nA.}
\end{figure}

In this section, we demonstrate the tuning of the circuit into the regime where the tunnelling current can exceed the critical current of the thin film. First, we tune the critical current, which depends on the physical parameters of the film such as width, thickness, and superconducting film composition. Moreover, it depends on external parameters such as the bias current (dc and rf) as well as the temperature and magnetic field~\cite{Annunziata_2010,S_Frasca,Luomahaara2014KineticIM}. Here, in order to reduce $I_C$ below the maximum tunnelling current, we use a 0.6~T out-of-plane magnetic field, achieving $I_C\sim33$~nA, as shown in Fig.~\ref{fig:2}(a). The normal state resistance is 75~k$\Omega$, predominantly dominated by the polysilicon layer above the thin film. We note that, in the future, a separate current bias tab could be added, enabling an independent current biasing system for the film and the SET.


Next, we bias the SET to demonstrate the required tunnelling currents can be achieved. Figure~\ref{fig:2}(b) shows an $I_D$ map versus gate-source and drain-source voltages that reveal the characteristics of Coulomb diamonds of the SET (blue diamond-shaped low conductance regions). Most importantly, we observe regions where the current level can exceed $I_C$. We show a particular example in Fig.~\ref{fig:2}(c), where we plot a trace at fixed $V_\text{DS}=-12$~mV showing a Coulomb oscillation with a peak current above $I_C$. 

Next, we demonstrate the capability to force the phase transition conditional to the current flowing through the SET. Figure~\ref{fig:2}(d) reveals preliminary results, plotting the change in reflected power as a function of drive frequency for several values of $I_D$, referenced with respect to the $I_D$ = 0~nA state. For this measurement, we set $V_{GS}=1$~V, well in the on state of the SET, and apply a $V_\text{DS}$ bias to set the current level through the film and SET. The plot shows two blue lines where the TiN film is still in the superconducting state. At the resonance frequency, $f_0=334$~MHz, the change is maximal. Notably, this change is enhanced when we further measure the change in reflected power at an applied drain current of $40$~nA, i.e., passed the phase transition (red curve). 

Irrespectively of the specific state of the resonator, we note that, as a consequence of the applied magnetic field, the resonator in the superconducting state became substantially undercoupled (with respect to its zero field state), thus reducing the variation of the reflection coefficient to $\approx 10$\%. As we shall see in Sec.~\ref{sec:3}, the circuit could be redesigned to have a larger $\beta$ in the superconducting state to optimize the sensitivity of the PTSET.  

In this Section, we have demonstrated the ability to meet the conditions required to link the phase transition of the superconducting thin film to the resistance state of the FET. Furthermore, we have observed the enhancement of the change in reflection coefficient once the superconducting-to-normal phase transition occurs. Next, we benchmark the enhancement in sensitivity when operating in PTSET mode.

\section{\label{sec:2}PTSET sensitivity benchmark}

\begin{figure}[ht!]
\includegraphics[width=\columnwidth]{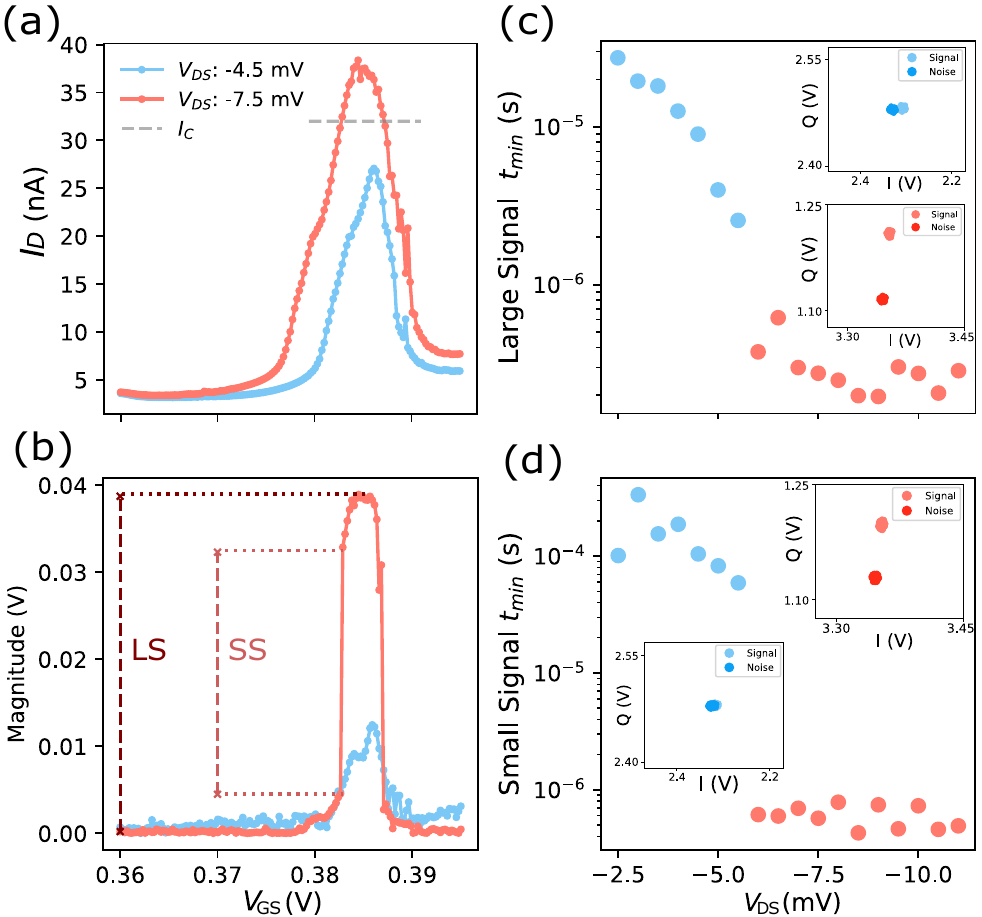}
\caption{\label{fig:3} \textbf{PT-SET SNR characterization.} 
(a) Coulomb blockade oscillation plot: Drain current as a function of the applied gate voltage for -7.5~mV (red) and -4.5~mV( blue) DS bias. (b) Magnitude of the reflected rf voltage as a function of applied gate-source voltage for -7.5~mV (red) and -4.5~mV (blue) DS bias cases. The vertical dashed lines represent our definition of large and small signal regimes, where the charge sensing is mimicked by a $V_\text{GS}$ shift larger or smaller than the linewidth of the Coulomb peak, respectively. For the small signal regime, we take points separated by $V_\text{GS}=467$~$\mu$V. (c,d) Minimum integration time versus DS voltage in the large signal and small signal regime, respectively. The insets show the Fresnel lollipops for signal and noise in the large signal regime for both -7.5~mV (red) and -4.5~mV (blue) in DS bias.} 
\end{figure}

Figures~\ref{fig:3}(a) and (b) present, respectively, dc and rf transfer curves for different source-to-drain biases: $V_{DS} = -4.5$~mV, where no superconducting-to-normal transition occurs for any gate voltage (blue traces), and $V_{DS} = -7.5$~mV, where the phase transition may be triggered (red traces). Focusing on Fig.~\ref{fig:3}(b), where we plot the magnitude of the reflected rf voltage as a function of $V_\text{GS}$, we observe how the blue trace tracks the $I_D-V_\text{GS}$ curve of the device as reported in panel (a). However, in the case of the red trace, we observe a sudden jump in reflected voltage once $I_D=I_C$, i.e. the condition for the phase transition. In the following, we exploit this abrupt voltage change to demonstrate enhanced sensitivity.

To benchmark the sensitivity of the PTSET, we use the metric of the minimum integration time ($t_{\mathrm{min}}$), i.e. the integration time ($t_{\mathrm{int}}$) needed to resolve, with a power signal-to-noise ratio (SNR) of 1, a charge sensing event ~\cite{10.1063/5.0088229}: 

\begin{equation}
    t_\text{min}=t_\text{int}/\text{SNR}  
\label{eq:SNR}
\end{equation}

Furthermore, we benchmark the device for both the large signal (LS) and small signal (SS) regimes -- mimicking a charge sensing event that produces a $V_\text{GS}$ shift much larger or smaller than the peak line-width respectively, the latter being the most commonly observed regime in spin qubit readout experiments. In the LS regime, we mimic the two charge states by gate biases at the top and bottom of the Coulomb peak. In the SS regime, we use gate biases in the region of steepest gradient (see Fig.~\ref{fig:3}(b) for a graphical depiction of the response in the LS(SS) regimes in the PTSET mode).


To extract the SNR, we acquire time traces of the in-phase (I) and quadrature (Q) components of the reflected voltage which we plot in the insets of Fig.~\ref{fig:3}(c,d) for a fixed integration time of 637~$\mu$s. We do so for the two gate bias conditions creating two separated data blobs (signal and noise). Then we calculate the SNR following the expression,


\begin{equation}
    \mathrm{SNR} = \frac{\left(I_{\mathrm{0}}-I_{\mathrm{1}}\right)^{2}+\left(Q_{\mathrm{0}}-Q_{\mathrm{1}}\right)^{2}}{0.25\left(\sigma_{\mathrm{0}}+\sigma_{\mathrm{1}}\right)^{2}},  
\label{eq:SNR}
\end{equation}

\noindent where $I_{\mathrm{0}}$, $I_{\mathrm{1}}$, $Q_{\mathrm{0}}$ and $Q_{\mathrm{1}}$ represent the coordinates in the IQ plane of the centres of the two data blobs and $\sigma_{\mathrm{0,1}}$ represent their corresponding 2D standard deviations, hence the denominator corresponding to their average across the two blobs~\cite{PhysRevApplied.21.044016}.  

In Fig.~\ref{fig:3}(c) and (d), we can see the resulting  $t_\mathrm{min}$ plotted as a function of the applied DS bias for both the LS and SS regimes, respectively. In Fig.~\ref{fig:3}(c), the LS regime, $t_\mathrm{min}$ exhibits a gradual decrease as a function of DS voltage followed by a sharp reduction for $V_\text{DS}\lesssim-6$~mV (the minimum $|V_\text{DS}|$ that creates the condition $I_D=I_C$). The gradual decrease can be explained by the expected improvement of the rfSET sensitivity with increasing current~\cite{PhysRevX.13.011023}. However, the previously mentioned abrupt decrease, corresponding to nearly an order-of-magnitude improvement in the sensitivity of the sensor, is attributed to the phase transition, i.e. the sensor being operated in PTSET mode. Beyond the phase transition bias, we observe a saturation of $t_\mathrm{min}$. The SNR enhancement can be clearly appreciated in the insets, where we compare the sensor response in the IQ plane in the normal SET mode (blue blobs acquired at $V_\text{DS}=-4.5$ mV) and PTSET mode (red blobs acquired at $V_\text{DS}=-7.5$ mV).

In the SS regime, panel (d), $t_{\mathrm{min}}$ displays a qualitatively similar behaviour although with a larger signal enhancement for the PTSET mode. For $V_\text{DS}\lesssim-6$~mV, $t_{\mathrm{min}}$ decreases abruptly, displaying a sensitivity enhancement of nearly two orders of magnitude. Again, after the sharp decrease, $t_\mathrm{min}$ saturates. While the insets in the SS regime exhibit the same qualitative behaviour as those in the LS regime, the quantitative change in state separation is larger, highlighting the substantial sensitivity enhancement in this most common regime for spin qubit readout experiments. Of minor importance but, we point out that the smaller rate of reduction of $t_{\mathrm{min}}$ with $V_\text{DS}$ in the normal SET regime can be explained as a consequence of the competing effects of increased current and reduced transconductance (wider Coulomb peak) as $V_\text{DS}$ is increased.



\section{\label{sec:3} PTSET optimization}



Finally, we address the question of which PTSET design parameters best maximize the sensor’s charge sensitivity. We address the problem using the Quite Universal Circuit Simulator (QUCS) software package~\cite{QUCS_Brinson} to design and simulate the PTSET. In our model, we utilize the experimental values of the circuit components and optimize the coupling coefficient in the superconducting state (via the choice of $C_C$). 



\begin{figure}[ht!]
\includegraphics[width=\columnwidth]{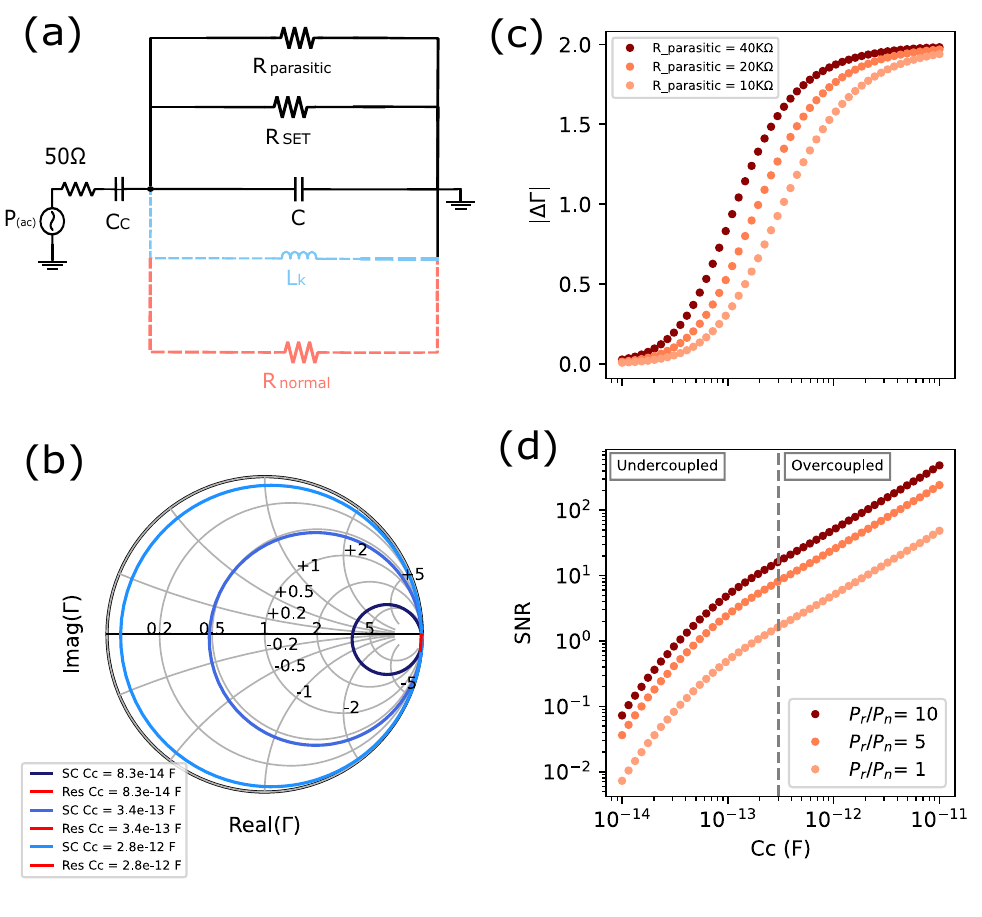}\caption{\label{fig:4} \textbf{Theoretical SNR limits} (a) The schematic model of the matching circuit with both TiN behaviour possibilities: resistive state (red), superconducting state (blue).  (b) Smith chart for S parameter simulations in both resistive and superconducting models, for three different $C_C $ values.  (c) Reflection coefficient magnitude change from superconducting to resistive state as a function of  $C_C$ , for three different $R_\mathrm{parasitic} $ values. (d) SNR as a function of  $C_C$ for several $P_\mathrm{r}/P_\mathrm{noise}$ ratios and $R_\text{parasitic}=20$~k$\Omega$. The vertical line marks the $C_C $ value in which the circuit is critically coupled.}
\end{figure}

Figure~\ref{fig:4}(a) shows the schematic of the simulated circuit. All lumped element values are known except for the circuit's parasitic resistance ($R_\text{parasitic}$). The parasitic resistance lumps all internal losses of the resonator, accounting for effects such as contact series resistances and MOM capacitor dielectric losses. We extract a value of 20~ k$\Omega$ by fitting the rf circuit model to the reflection coefficient vs frequency data at zero magnetic field when the SET is in the "off" state. For the SET resistance in the "off" state ($R_\mathrm{SET}^\text{off}$), we assign a value of  $1 ~\mathrm{G}\Omega$ . We note that as long as $R_\mathrm{SET}^\text{off}\gg R_\text{parasitic}$, its influence on the simulated results is negligible. In the "on" state, we approximate $R_\mathrm{SET}=100~ k\Omega$, greater than twice the resistance quantum.


Next, we perform S-parameter simulations for both operating regimes of the PTSET: (i) the superconducting state, where we link the high resistance state of the SET with the superconducting state of the film ($R_{\mathrm{SET}}= 1$~G$\Omega$ and $L_\mathrm{k}=122$~nH, blue path) and (ii) the normal state, where we link the low resistance state of the SET with the normal conducting state of the film ($R_\mathrm{SET}=100$~k$\Omega$, $R_\mathrm{normal}$ = $75~ k\Omega$, red path). To evaluate the PTSET performance across the design space, we vary the coupling capacitance ($C_C$), our design variable of choice.  

Figure~\ref{fig:4}(b) illustrates the resonator response across three representative  $C_C$ values on a single Smith Chart. In the superconducting state (blue traces), we see constant resistance circles going from undercoupled in the low $C_C$ case (dark blue), to overcoupled in the large $C_C$ case. Notably, in the normal state (red curves), the behaviour of the resonator becomes approximately invariant to the changes in $C_C$ since the impedance of the circuit is mainly resistive (dominated by the parallel combination of $R_\text{parasitic}$ and $R_\text{normal}$) and much larger than the characteristic impedance, i.e. the circuit is undercoupled for all choices of $C_C$. 

We compute the variation in reflection at the resonant frequency between superconducting and normal states for each choice of $C_C$ and plot it in Fig.~\ref{fig:4}(c) for various values of $R_\mathrm{parasitic}$. We observe the trend anticipated in the Smith chart analysis. For low $C_C$, the system is undercoupled in the superconducting state, yielding a small contrast between the superconducting and normal states. As we increase $C_C$, and hence the coupling coefficient, |$\Delta\Gamma$| increases until it saturates to $|\Delta\Gamma|=2$, corresponding to the maximum possible change in reflection coefficient. Variations in $R_\mathrm{parasitic}$ generate a lateral shift, pushing the curve towards lower $C_C$ the larger $R_\text{parasitic}$, yet the value does not affect the general trend of the curve. However, the change in the reflection coefficient does not fully determine the maximum sensitivity of the PTSET. To calculate the achievable SNR of the sensor, we need to consider the dependence on the input power that can be applied to the resonator ($P_\text{in}$),

\begin{equation}
    \mathrm{SNR} = |\Delta\Gamma|^{2}\frac{P_\mathrm{in}}{P_\mathrm{noise}}.  
\label{eq:Signal}
\end{equation}

Here, $P_\text{noise}$ refers to the noise power, typically determined by the noise of the first stage of amplification. For this first discussion, we consider it constant for all bias conditions. We then link the power delivered to the resonator with that dissipated across the SET and parasitic resistor, $P_\text{r}$, 

\begin{equation}
    P_\mathrm{in} = P_\text{r} \frac{(1 + \beta)^{2}}{4\beta},  
\label{eq:Pin}
\end{equation}

\noindent where $P_\text{r}=v_\mathrm{SET}^2 (R_\mathrm{SET} // R_\text{parasitic})$ and $v_\mathrm{SET}$ is the voltage drop across the SET. We then calculate the SNR as a function of $\beta$ and $P_\text{r}$,

\begin{equation}
    \mathrm{SNR} = \frac{P_\mathrm{r}}{P_\mathrm{noise}} \frac{(1 + \beta)^{2}}{4\beta} |\Delta\Gamma|^{2}\rightarrow \frac{P_\mathrm{r}}{P_\mathrm{noise}}\beta.  
\label{eq:Signal}
\end{equation}

We plot the results in Fig.~\ref{fig:4}(d), where we present the SNR vs $C_C$ for different $P_\mathrm{r}/P_\mathrm{noise}$ ratios, or equivalently measurement bandwidths. By fixing this ratio, we also guarantee a constant voltage drop across the SET, hence avoiding overdriving the sensor for any coupling condition. For low $C_C $ values, the SNR tends to 0 for any input power since the change in the reflection coefficient does vanish. For increasingly larger $C_C$, the SNR ratio increases first quadratically and then linearly since $\beta \propto C_C^2/(C+C_C)$~\cite{Ahmed2018Radio-FrequencySensing}. Overall these results indicate the benefit of overcoupling the matching network when the PTSET is in the superconducting state. This feature can be understood by noting that, for larger coupling coefficients, the system can be driven more strongly without overdriving the SET. In the limit of very large $\beta$, the rf signal undergoes a change from soft total reflection ($0^\circ$ phase shift) in the superconducting state to hard total reflection ($180^\circ$ phase shift) in the normal state. Increasingly larger $P_\mathrm{r}/P_\mathrm{noise}$ ratios (or lower measurement bandwidth) result in larger SNR, but overall the curves followed the same trend. 

We put these results in perspective by making a comparison with the standard rfSET. For the case $R_\text{parasitic}=20$~k$\Omega$ used here, the change in reflection coefficient is small and proportional to $\Delta R_\text{eq}/R_\text{eq}$ where $R_\text{eq}$ is the parallel combination of $R_\text{parasitic}$ and $R_\text{SET}$ and  $\Delta R_\text{eq}$ is its change between the on and off states of the SET~\cite{Thomas2025}. This consideration corresponds to $|\Delta\Gamma|_\text{rfSET}=0.167$, much smaller than the maximum $|\Delta\Gamma|\rightarrow2$ achievable for the PTSET. In addition, the small $\Delta R_\text{eq}/R_\text{eq}$ for the rfSET case, results in an optimal driving condition at $\beta=1$ rather than at large $\beta$ for the PTSET case~\cite{10.1063/5.0088229}. Overall, the improvement in SNR can be expressed as, 

\begin{equation}
    \frac{\text{SNR}}{\text{SNR}_\text{rfSET}} =\frac{(1 + \beta)^{2}}{4\beta}\frac{|\Delta\Gamma|^{2}}{|\Delta\Gamma|^{2}_\text{rfSET}}\rightarrow \frac{\beta}{|\Delta\Gamma|^{2}_\text{rfSET}},  
\label{eq:SNR}
\end{equation}

\noindent where for $R_\text{parasitic}=20$~k$\Omega$ exposed here and for a well overcoupled case in PTSET mode ($\beta=10$), we find a theoretical SNR improvement of a factor of more than 350. The improvement in PTSET mode with respect to rfSET mode decreases as $\Delta R_\text{eq}/R_\text{eq}$ increases ($R_\text{parasitic}$ increases) eventually the ratio tending to 1 as $R_\text{parasitic}\rightarrow\infty$, but this is experimentally challenging.

\section{\label{sec:4}Conclusions}

We have demonstrated a proof-of-concept phase-transition single-electron transistor (PTSET) and introduced a circuit designed to enhance its response. Analytical modelling and numerical simulations establish the performance limits of the device and predict substantial improvements in signal-to-noise ratio compared with conventional rfSETs, highlighting the potential of phase-transition sensing for high-sensitivity charge detection.

Several aspects of the PTSET remain to be experimentally explored. In particular, its switching and recovery dynamics will determine the ultimate speed of the detector. Although these times could not be resolved with the present experimental setup, the electrical switching time is expected to be limited by the inductive time constant, which is on the order of picoseconds for the circuit investigated here. The recovery to the superconducting state is instead expected to be governed by thermal relaxation. Recovery times of a few to tens of nanoseconds have been demonstrated in NbTiN SNSPDs \cite{8118177,Miki_2009}. Given the lower superconducting energy gap of TiN, we estimate recovery times in the range of approximately 10–100 ns for our films, suggesting that the PTSET could support the timescales required for semiconductor spin-qubit readout. 

More broadly, our results establish phase-transition sensing as a promising route towards high-sensitivity, fully integrated charge detection. The use of a superconducting transition to amplify the response of a charge sensor could enable low-power readout architectures for scalable semiconductor quantum processors. The same principle may also be extended to other cryogenic sensing applications, including single-photon detection and the development of fully integrated superconducting photon detectors.

\section{\label{sec:Methods} Appendix}
\subsection{\label{sec:Methods:RF-Reflecto} Appendix A: Critical Current Tuning} 

For a configuration where no magnetic field is applied and the temperature is below 100 mK, the critical switching current (the current level in which the superconductor changes from the state to the resistive state) of the inductive thin film (for the geometry used) is ~$\sim1~\mathrm{\mu}$A ~\cite{Swift:2025wkk}, which is higher than the current level achievable for standard Coulomb blockade devices~\cite{RevModPhys.79.1217}. To reduce the critical current, we review the possible parameters to bring our setup into the desired configuration, this is a critical current below the maximum tunnelling current, $I_C<I_D^\text{max}$. We comment on different tuning strategies to achieve PTSET operation.  

The first option is to pass an independent bias current to the inductor such as the thin film approaches $I_C$ independently of the tunnelling current. This option is appealing since it provides an all-electrical tuning knob. However, the current circuit design, see Fig.~\ref{fig:1}(a), does not support an independent bias current but future generations may incorporate such current bias electrode. 

The second option is to use temperature tuning. The theoretical temperature-dependence of critical current can be derived using BCS (Bardeen-Cooper-Schrieffer) theory and is called the Bardeen formula~ \cite{RevModPhys.34.667}, demonstrating how the critical current reduces as a function of temperature: 

\begin{equation}
    I_{\mathrm{C}}(T) = I_{\mathrm{C}}(0) \left(1-\left[\frac{T}{T_{\mathrm{c}}}\right]^2\right)^\frac{3}{2},
\label{eq:Lk_T_dependence}
\end{equation}

\noindent where $I_C (T))$ represents the critical current at a certain temperature, $T$ is the temperature of the system, and $T_C$ is the critical temperature of the semiconductor.
The critical temperature of these inductive films is around 1.1~K. Using the global temperature as tuning parameter comes with disadvantages since it also reduces the sensitivity of the sensor as the temperature is increased. However, it is possible to implement local heating techniques, allowing independent temperature bias of the thin film and the SET. 

The third option, the one used in this study, utilizes magnetic field tuning. The critical current similarly decreases with an applied magnetic field (in this case perpendicular to the thin film) with a functional dependence given by the Kim-Type formula \cite{PhysRev.129.528}:

\begin{equation}
    I_C(B_\perp) = \frac{I_C(B_\perp = 0)}{1 + \frac{B_\perp}{B_0}},
\label{eq:Lk_T_dependence}
\end{equation}

\noindent where $I_C(B_\perp)$ is the critical current for a determined perpendicular magnetic field, and $B_0 $ is the critical magnetic field. In this work, we utilized the magnetic field to tune the critical current. However, we note that the superconductor's inductance increases with the applied magnetic field. Specifically, a strong out-of-plane field reduces the Cooper-pair density, increasing the kinetic inductance ($L_{\mathrm{K}}$) and thereby weakening the coupling to the reflectometry circuit. In this strongly undercoupled regime, the change in reflection coefficient becomes correspondingly small \cite{PhysRevApplied.5.044004}. 

\subsection{\label{sec:Methods: Retrapping current} Appendix B: Retrapping current}

Notably, $I_C$ represents the switching current required to transition the device from the superconducting to the normal state. The retrapping current (the current required for transitioning back from normal to superconducting) observed under identical conditions was $I_\mathrm{re}\sim22$~nA  . This current is directly related with the device recovery time and dynamics which have been mentioned previously.

\section{Acknowledgement}
We thank Charan Kocherlakota, Debargha Dutta, and Jonathan Warren of Quantum Motion for their technical support. A.G.-S. acknowledges an Industrial Fellowship from the Royal Commission for the Exhibition of 1851. M.F.G.Z. acknowledges a UKRI Future Leaders Fellowship [MR/V023284/1].

\section{Author Contributions}

G. A. I. and T. H. S. analysed the data. M. F. G. Z. and F.E.v.H. conceived the experiment. A. G. S. conceived and designed the superconducting structures. F. O. designed the integrated circuits. M. F. G. Z. supervised the work. All authors contributed to the writing of this manuscript.


\section{Competing interests}

M.F.G-Z and F.E.v.H. are inventors of a relevant patent (WO2026093133A1).


\bibliography{General}

\end{document}